\documentclass[aps,pra]{revtex4}
\usepackage{latexsym}
\usepackage{epsfig}
\newcommand{\beq}{\begin{eqnarray}}
\newcommand{\eeq}{\end{eqnarray}}
\newcommand{\be}{\begin{eqnarray}}
\newcommand{\ee}{\end{eqnarray}}
\newcommand{\bk}{{\bf k}}

\newcommand{\bx}{{\bf x}}

\begin{document}

\title{Growth of perturbations in dark matter coupled with quintessence}
\date{\today}

\author{Tomi Koivisto}
\email{tomikoiv@pcu.helsinki.fi}
\affiliation{Helsinki Institute of Physics,FIN-00014 Helsinki, Finland}
%\affiliation{Department of Physical Sciences, University of Helsinki, FIN-00014 Helsinki, Finland}
\affiliation{Department of Physics, University of Oslo, N-0316 Oslo, Norway}
%\author{Hannu Kurki-Suonio}
%\email{hkurkisu@pcu.helsinki.fi}
%\author{Finn Ravndal}
%\email{finn.ravndal@fys.uio.no}
%\affiliation{Department of Physics, University of Oslo, N-0316 Oslo, Norway}

\begin{abstract}

We consider the evolution of linear perturbations in models with a nonminimal coupling between dark matter and scalar field dark energy. 
Growth of matter inhomogeneities in two examples of such models proposed in the literature are investigated in detail. Both of these models are based 
on a low-energy limit of effective string theory action, and have been previously shown to naturally lead to late acceleration of the universe. However, 
we find that these models can be ruled out by taking properly into account the impact of the scalar field coupling on the formation of structure in 
the dark matter density. In particular, when the transition to acceleration in these models begins, the interaction with dark 
energy enchances the small scale clustering in dark matter much too strongly. We discuss also the role of an effective small scale sound speed 
squared and the issue of adiabatic initial conditions in models with a coupled dark sector. 

\end{abstract}

\maketitle

\section{Introduction}

Cosmological observations suggest that about ninetysix percent of the present energy density in the universe resides in forms unknown to the standard 
model of particle physics. Measurements of the fluctuations in the cosmic microwave background (CMB) radiation are consistent with a nearly critical 
density universe\cite{Spergel:2003cb}, as predicted by the early inflation. Assuming the validity of general relativity, roughly two thirds of this 
density nowadays must be due to a negative-pressure component called dark energy\cite{Peebles:2002gy} (DE) to explain the acceleration of the universe 
indicated by distance-redshift curves of the high redshift supernovae\cite{Riess:2004nr}. The rest of the density is in gravitationally clustered 
matter with negligible pressure on cosmological scales. However, only a fraction of this matter consists of the (partly luminous) baryonic matter 
known to the standard model. Dark matter (DM) is needed for consistency with a convincing number of astrophysical observations, for example the flat 
rotation curves of galaxies\cite{Bertone:2004pz}. 

It has been proposed that the two dark components do not evolve independently but are non-minimally coupled. Due to their different pressures, 
the densities of these components decay with different rates. This rises the disturbing question why they appear to be of the same order of 
magnitude right now. An explicit coupling between DE and DM\cite{Wetterich:1994bg} could provide a phenomenological solution to this so called 
''cosmic coincidence problem''\cite{Zimdahl:2001ar,Chimento:2003ie,
Bento:2004uh,Cai:2004dk}. On the other 
hand, there is theoretical motivation\cite{Farrar:2003uw} to couple 
quintessence fields\cite{Zlatev:1998tr,Caldwell:1997ii} to DM (see, 
however\cite{Doran:2002bc}). In particular, if 
it is held that scalar-tensor theories should be described in the Einstein frame, it follows that
scalar fields with a nonminimal coupling to other matter are general predictions of all dimensionally reduced 
theories\cite{Albrecht:2001xt,Brax:2004ym}. In the context of string 
theory, the dilaton could be such a field, and indeed has been proposed as a candidate 
for DE \cite{Gasperini:2001pc, Bean:2000zm}. Since generally 
the dilaton coupling to different matter species could be different\cite{Garcia-Bellido:1992de} and such couplings to baryonic matter are tightly 
constrained by high precision solar system tests of general relativity\cite{Damour:1994zq}, it is of interest to concentrate on simpler models where 
the dilaton is only coupled to DM, while the possible coupling to baryons is assumed negligible.

This paper concerns the linear perturbations in such models. The purpose is to find how the formation of cosmological structure 
in pressureless, dust-like matter is influenced by a coupling to a scalar field, when this coupling is significant already at the background order as 
required if it is to address the coincidence problem. First the cosmological field equations for scalar field interacting with DM
are derived to linear order in section II. These equations have been presented in the literature before (for example, see \cite{Amendola:2003wa}), and 
the reader familiar with them may skip this section. However, there we hope also to somewhat clarify the relations of different possible 
formulations and theoretical frameworks for the coupled dark sector. In the two following sections we investigate  
specific cases. The simplest models of interacting DM and DE can be excluded by comparison to the CMB observations\cite{Hoffman:2003ru}, but more 
sophisticated scenarios exist\cite{Gasperini:2001pc, Bean:2000zm,Farrar:2003uw}. In 
section III we study a model featuring two DM components, only the 
subdominant of them interacting with quintessence\cite{Huey:2004qv}. A model where the DE-DM coupling is given by a polynomial function of the DE 
field\cite{Bean:2000zm} is considered in section IV. These two models are based on a low-energy limit of effective string theory action and have been 
previously shown to naturally lead to a late-time acceleration of the universe. Though the details of the mechanisms are different, the DE 
domination in each case is triggered by the time-evolving coupling. Large scale structure in both of the two scenarios have previously been claimed 
to fit observations as well as or better than the standard $\Lambda$CDM model. However, here a closer inspection shows that these models are ruled out 
because of the excessive DM structure they predict in the late-time universe. In subsection IVB we consider the adiabatic initial conditions in for a 
scalar field coupled to DM. In section V we discuss the relation of our results to other models and conclude. 

\section{Equations}

\subsection{Field equations and Bianchi identities}

We consider an action of the form

\be \label{action}
    \mathcal{S} = \int d^4x \sqrt{-g}[\frac{\kappa^2}{2}R - \frac{1}{2}\partial^\mu\phi\partial_\mu\phi - V(\phi) 
                  + \gamma_\chi (\phi)\mathcal{L}^{(\chi )} + \mathcal{L}^{(m)}],
\ee
where $\chi$ stands for an interacting matter species, and $m$ for non-interacting ones. The constant $\kappa \equiv (8 \pi G)^{-1/2}$.
We use the definition
\be 
    T_{\mu\nu}^{(i)} \equiv -\frac{2}{\sqrt{-g}}\frac{\delta (\sqrt{-g}\mathcal{L}_i)}{\delta(g^{\mu\nu})},
\ee
and the Lagrangian of the field $\phi$ is taken to consist of the two first terms in the action integral Eq.(\ref{action}). The Einstein equations 
resulting from this action can be written as
\be
    G_{\mu\nu} = \frac{1}{\kappa^2} (T_{\mu\nu}^{(\phi)} + \gamma_\chi(\phi)T_{\mu\nu}^{(\chi)} + T_{\mu\nu}^{(m)}),
\ee
and the equation of motion for the field $\phi$ is 
\be \label{eom}
    \Box \phi = V'(\phi) - \gamma_\chi'(\phi)\mathcal{L}^{(\chi)}.
\ee
The Bianchi identities then imply 
\be
    \nabla_\mu {T^{\mu}_\nu}^{(m)}=0,
\ee
\be \label{bianchic}
    \nabla_\mu {T^{\mu}_\nu}^{(\chi)}=(\delta^\mu_\nu\mathcal{L}^{(\chi)} - 
               {T^{\mu}_\nu}^{(\chi)})\frac{\gamma_\chi'(\phi)}{\gamma_\chi(\phi)}\nabla_\mu\phi.
\ee
Thus the energy momentum tensor for the interacting components are not separately conserved.

From now on we will assume that 
\be \label{lagran}
\mathcal{L}^{(\chi)} = -\sum_a m_a\delta (x-x_a(t))\frac{\sqrt{-g_{\mu\nu}\dot{x}^\mu\dot{x}^\nu}}{\sqrt{-g}},
\ee
where $|\dot{x}^i| \ll 1$, i.e. the interacting component behaves as dust. This is in fact the case most commonly studied in cosmology (for models 
with 
DE coupled to neutrinos, see Refs.\cite{Peccei:2004sz,Brookfield:2005td}). One notes that the term $\gamma_\chi(\phi)\mathcal{L}^{(\chi)}$ in the 
action 
Eq.(\ref{action}) can be naturally interpreted as a Lagrangian for 
particles with field-dependent masses, $m_a(\phi) = m_a\gamma_\chi(\phi)$. Such a Lagrangian can follow from a conformal 
transformation. Consider a scalar-tensor theory defined by a gravitational Lagrangian proportional to $R/\sqrt{\gamma(\phi)}$. Then 
the metric in the Einstein frame is $\hat{g}_{\mu\nu} \equiv g_{\mu\nu}/\sqrt{\gamma(\phi)}$. In terms of this metric, the action 
for the Lagrangian Eq.(\ref{lagran}) reads 
\be
S_\chi = \int d^4x \sqrt{-\hat{g}} \gamma(\phi)\hat{\mathcal{L}}^{(\chi)}.
\ee
(At this point one may have to redefine the field in order to have the canonical kinetic term for the scalar field as in the action 
Eq.(\ref{action}).) A more standard way, however, to describe scalar-tensor theories in the Einstein frame is to define the 
component 
\be \label{x-comp}
T_{\mu\nu}^{(X)} = \gamma_\chi(\phi)T^{(\chi)}_{\mu\nu}.
\ee 
This energy momentum tensor then depends on the field $\phi$, which turns out to be coupled to the trace $T^{(X)}$. For some 
purposes the $\chi$-formulation is more convenient, since there the divergence in Eq.(\ref{bianchic}) vanishes except for the
spatial part in the linear order. However, we will also use the $X$-formulation for some calculations, especially in subsection IIIB. 
With the variable mass particle interpretation, the difference is simply that $\rho_\chi$ is proportional to the number density of the DM
matter particles, while $\rho_X$ is their gravitating energy density. The fact that the action Eq.(\ref{action}) is equivalent to the 
case of a trace-coupled scalar field and the possibility to make direct contact with conformally transformed scalar tensor and variable mass 
particle theories are independent of the form of the coupling. However, it is required that the Lagrangian is effectively of the form 
Eq.(\ref{lagran}). From that follows (when $|\dot{x}^i| \ll 1$)
\be
\mathcal{L}^{(\chi)}=T^{(\chi)}=-\rho_\chi.
\ee 
These simple identities would not, of course, hold for a more general Lagrangian.

\subsection{Cosmological background and perturbations}

We consider a flat FRW universe. Including small perturbations about the background, we can write the line element as 
\be
      d s^2 = a^2(\tau)[-d\tau^2 + (\delta_{ij} + h_{ij}) dx^i dx^j].
\ee
This parameterisation of the metric perturbations is called the synchronous gauge. There we can characterize the scalar modes of $h_{ij}$ in the 
Fourier space as 
\be
       h_{ij}(\bx,\tau) = \int d^3k e^{i\bk \cdot \bx} \biggl[\hat{k}_i\hat{k}_jh(\bk,\tau) \mbox{} +  (\hat{k}_i\hat{k}_j - 
                          \frac{1}{3}\delta_{ij})6\eta(\bk,\tau)\biggr],
\ee
when $\bk=k\hat{k}$. In the following, prime is a derivative with respect to $\phi$, an overdot represents derivative with respect to conformal time 
$\tau$, 
and $H$ 
is the conformal Hubble parameter $H \equiv \dot{a}/a$. These and mostly all other notations and conventions are those of Ma and 
Bertschinger\cite{Ma:1995ey}. With these, the Friedmann equation reads  
\be \label{friedmann}
H^2 = \frac{a^2}{3\kappa^2}\left( \frac{\dot{\phi}^2}{2a^2} + V(\phi) + \gamma_\chi \rho_\chi + \rho_m \right), 
\ee
while evolution of the scalar field is governed by the equation 
\be \label{eom2}
\ddot{\phi} + 2H\dot{\phi} + a^2 V(\phi)' = -a^2\gamma_\chi'(\phi)\rho_\chi,
\ee
following from Eq.(\ref{eom}). In addition to the interacting species and the scalar field, we include standard CDM, baryons and radiation in the 
universe. Then the energy densities scale as usually, and therefore, with the scale factor normalized to unity today, we may write
\be
\rho_\chi=\frac{\rho_\chi^0}{a^3}, \qquad 
\rho_c=\frac{\rho_c^0}{a^3}, \qquad 
\rho_b=\frac{\rho_b^0}{a^3}, \qquad 
\rho_r=\frac{\rho_r^0}{a^4}.
\ee
Here superscript $0$ denotes the density today. However, instead of $\rho_\chi$ the gravitating energy density is now the density $\rho_X = 
\gamma_\chi \rho_\chi$ from definition Eq.(\ref{x-comp}), as implied by Eq.(\ref{friedmann}).  

The equations for the perturbed metric are
\be \label{emetric1}
k^2\eta - \frac{1}{2}H\dot{h} =
\frac{a^2}{2\kappa^2}\delta T^0_0,
\ee
\be 
k^2\dot{\eta}=\frac{a^2}{2\kappa^2}ik_i\delta T^0_i,
\ee
\be \label{emetric3}
\ddot{h}+2H\dot{h}-2k^2\eta=-\frac{a^2}{\kappa^2}\delta T^i_i,
\ee
\be
\ddot{h}+6\ddot{\eta}+2H(h'+6\dot{\eta})-2k^2\eta=\frac{3a^2}{\kappa^2}(\hat{k}_i\hat{k}_j-\frac{1}{3}\delta_{ij})\Sigma^i_j,
\ee 
where the components of the perturbed stress energy tensor are
\be 
\delta T^0_0 = -\rho_r\delta_r- \rho_\chi(\gamma_\chi'\varphi+\gamma_\chi\delta_\chi) -
   \rho_b \delta_b  -
   \rho_c \delta_c  - \frac{1}{a^2}\dot{\phi}\dot{\varphi} - V'\varphi,
\ee
\be
ik_i\delta T^0_i=\frac{4}{3}\rho_r\delta_r\theta_r+
\gamma_\chi \rho_\chi\theta_\chi +
\rho_b\theta_b +
\dot{\phi}k^2\dot{\varphi},
\ee
\be
\delta T^i_i = \rho_r\delta_r + 3c_b^2\rho_b\delta_b + \frac{3}{a^2}\dot{\phi}\dot{\varphi} 
       - 3 V'\varphi.
\ee
We have grouped together photons and 
neutrinos with subscript $r$, since their contribution is as usual, and used $\varphi$ to denote $\delta \phi/\phi$ as in\cite{Ferreira:1997hj}. 
In addition to the field equations, we need the equations of motion for the individual components. For the perturbations of the coupled scalar field, 
we derive the Klein-Gordon equation from Eq.(\ref{eom}) to find 
\be \label{klein}
\ddot{\varphi} + 2H\dot{\varphi} + (a^2 V'' + k^2)\varphi + \frac{1}{2}\dot{\phi}\dot{h} = 
-a^2\rho_\chi(\gamma_\chi'\delta_\chi+\gamma_\chi''\varphi).
\ee 
For the $\chi$-component, the Bianchi identities Eq.(\ref{bianchic}) in the linear order are
\be 
\dot{\delta}_\chi = -\theta_\chi - \frac{1}{2}\dot{h},
\ee
\be \label{vel}
\dot{\theta}_\chi = -H\theta_\chi + \frac{\gamma_\chi'}{\gamma_\chi}(k^2\varphi - \dot{\phi}\theta_\chi).
\ee
Conservation equations for uncoupled matter stay unmodified from the ones given in\cite{Ma:1995ey}.

\section{Model with two cold dark matter species}

\subsection{Background solution}

In this section we investigate the model proposed by Huey and Wandelt\cite{Huey:2004qv}. In this model quintessence with an exponential potential,
\be
V(\phi) = V_0 e^{-\kappa\alpha\phi},
\ee
is exponentially coupled to $\chi$-matter 
\be \label{coupling}
\gamma_\chi(\phi) = e^{\kappa\beta(\phi-\phi_c)}.   
\ee
There is also present an uncoupled species of cold DM. At early times, the field behaves as standard exponential 
quintessence, but when it reaches the minimum of its effective potential, $V_{eff} = V + \gamma_\chi\rho_\chi$, it evolves to a 
potential-dominated stage driving the acceleration of the universe.   

There is motivation for the additional complexity of introducing two kinds of DM. It is well known that with minimally coupled exponential 
quintessence the tracking solutions do not yield an accelerating phase\cite{Copeland:1997et}. If the field is exponentially coupled to matter, there 
exists an accelerating attractor\cite{Amendola:1999er}, but obviously the field could not have reached it until today lest there would not have been 
a matter dominated era at all. Moreover, it is known that in the such models the coupling between DE and DM is tightly constrained by the CMB 
spectrum\cite{Amendola:2002bs} and other cosmological data\cite{Franca:2003zg}. 
%One might think that when only a fraction of CDM is coupled to a 
%scalar field, the situation is at least alleviated. However, we will see that this not the case for this particular model.

\begin{figure}[h]
\begin{center}
\includegraphics[width=0.7\textwidth]{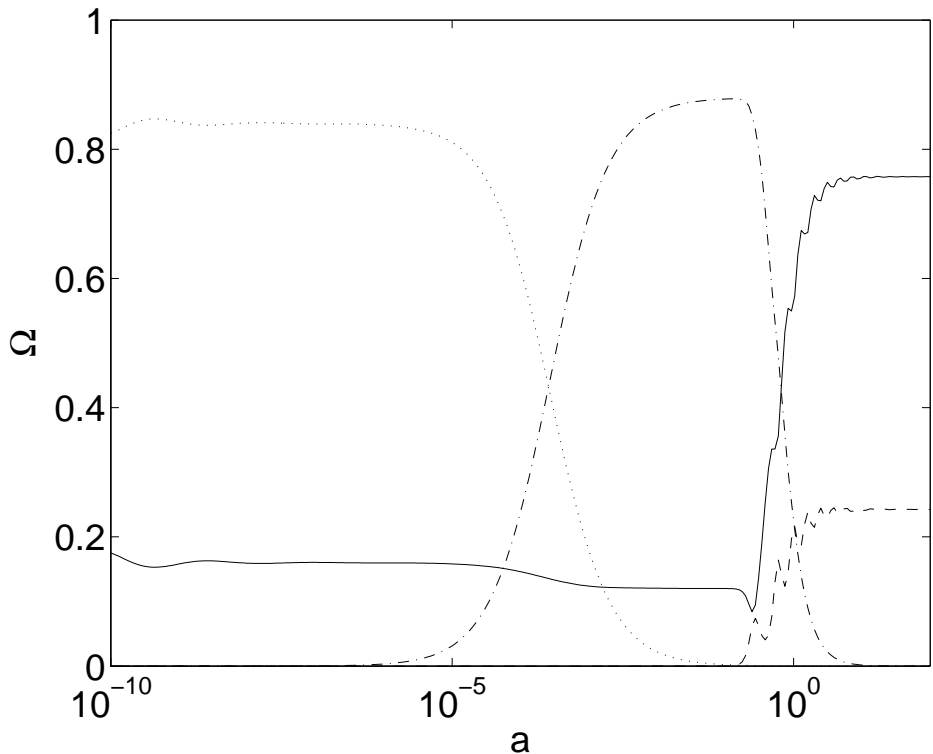} 
\caption{\label{omegas_i}Evolution of the fractional energy densities in model of section III. The solid line is $\Omega_\phi$, the dashed line 
$\Omega_X$, the dotted line is $\Omega_r$ and the dash-dotted line $(\Omega_c+\Omega_b)$.}
\end{center}
\end{figure}

The amount of quintessence during BBN must be small, constraining $\alpha \gtrsim 5$ \cite{Copeland:1997et}. The onset of acceleration should occur 
after matter-radiation equality, which is achieved when $\kappa\phi_c > 108/\alpha$. Finally, consistent late-time background evolution requires 
$\alpha/\beta \lesssim 3/5$ \cite{Huey:2004qv}. As a representative parameter combination we will use $\alpha=5$, $\beta=15$, $\phi_c=25$ in numerical 
calculations. Fig.(\ref{omegas_i}) shows then the evolution of the fractional energy densities
$\Omega_i \equiv \kappa^2 \rho_i/(3H^2)$. We have arranged the interacting and standard CDM 
components to be of equal importance today. We see some transient oscillations at the late epoch when the coupling term begins 
to hold the field at the minimum of its effective potential. However, asymptotically the coupling will keep the ratio of quintessence and interacting 
DM densities constant and the field evolution is again characterized with a constant scaling.  

\begin{figure}[h]
\begin{center}
\includegraphics[width=0.45\textwidth]{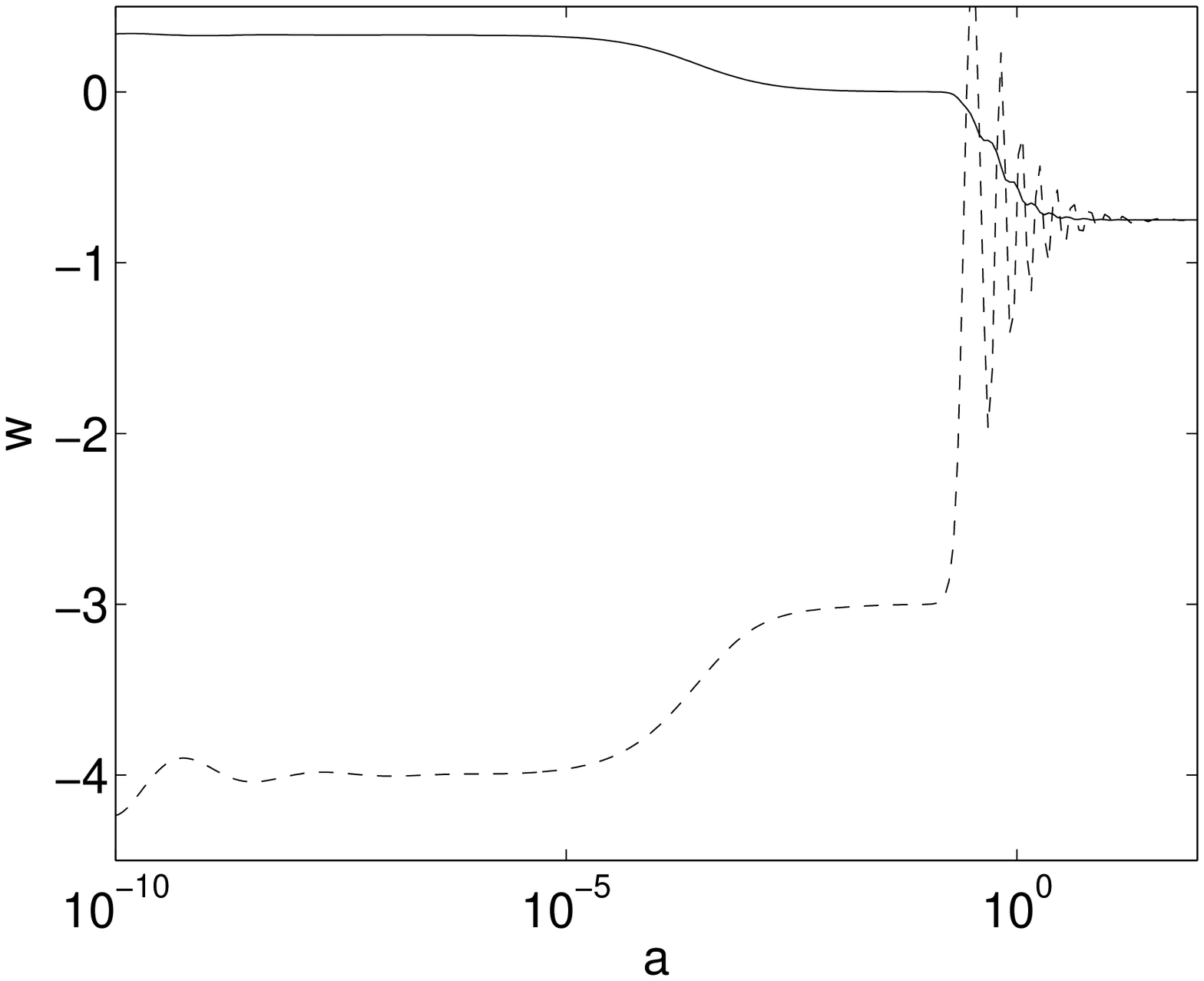} 
\includegraphics[width=0.45\textwidth]{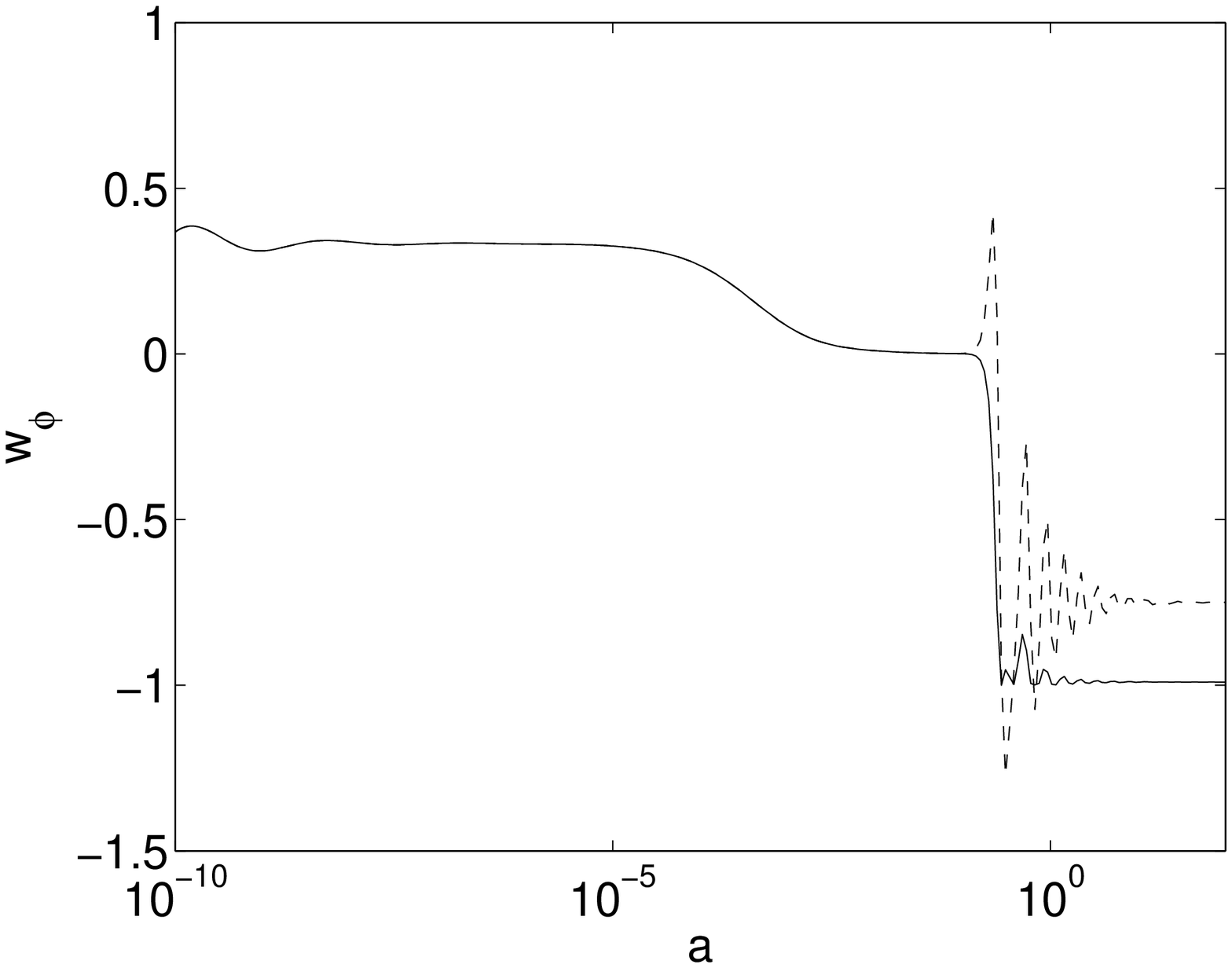} 
\caption{\label{eos_i}Left panel: Total equation of state $w$ (solid line) and the effective $w_X^{(eff)}$ (dashed line) for interacting CDM in the 
model of section III. Since $w<-1/3$ today, the universe accelerates. Right panel: Equation of state for the quintessence in the same model. 
Solid line is the formal $w_\phi$, and dashed line is the effective $w_\phi^{(eff)}$.}
\end{center}
\end{figure}

Consider the component $\rho_X \equiv \gamma_\chi\rho_\chi$. Since during the tracking phase, assuming a constant background equation of state $w_B$, 
\be \label{tracker}
\kappa\phi = \frac{3(1+w_B)}{\alpha}\log(a) + constant,
\ee
then $\rho_X \sim a^{3[-1+\frac{\beta}{\alpha}(1+w_B)]}$. Thus this energy density is growing in a phantom-like manner. Its effective equation-of 
state parameter, defined by
\be \label{wxeff}
\dot{\rho}_X + 3H\rho_X(1+w_X^{(eff)}) = 0,
\ee
is shown together with the total equation of state of the universe in the left panel of Fig.(\ref{eos_i}). Similarly we 
define the effective equation of state for the field $\phi$:
\be \label{wpeff1}
\dot{\rho}_\phi + 3H\rho_\phi(1+w_\phi^{(eff)}) = 0.
\ee
Then we have from Eq.(\ref{eom2}) that
\be \label{wpeff}
\dot{\rho}_\phi + 3H(\rho_\phi+p_\phi) = -\gamma_\chi'(\phi)\rho_\chi\dot{\phi} \quad \Rightarrow \quad 
       w_\phi^{(eff)} = \frac{p_\phi+\gamma_\chi'\rho_\chi\dot{\phi}/(3H)}{\rho_\phi}.
\ee
We plot both $w_\phi$ and $w_\phi^{(eff)}$ in the right panel of Fig.(\ref{eos_i}). 


\subsection{Perturbation evolution}

Defining $C \equiv \log(\gamma_\chi)$, the coupled perturbation equations for the component $X$ can be written as\cite{Amendola:2003wa}
\be \label{edelta}
\dot{\delta}_X = -\theta_X - \frac{1}{2}\dot{h} + C'\dot{\varphi}+C''\dot{\phi}\varphi,
\ee
\be \label{etheta}
\dot{\theta}_X = -H\theta_X + C'(k^2\varphi - \dot{\phi}\theta_X).
\ee
For the largest scales the source term in Eq.(\ref{etheta}) can be neglected, and thus no velocity perturbations are generated. Since then
$\dot{\delta_c}-\dot{\delta}_X = C'\dot{\varphi}$, the coupling has a small effect on $\delta_X$, which then follows closely the standard CDM 
overdensities. 

To find the small scale evolution of perturbations, we follow 
derivations found in Refs.\cite{Amendola:1999er,Amendola:2001rc,
Amendola:2002mp,Amendola:2003wa}. Using a linear combination of Eqs.(\ref{emetric1}) and (\ref{emetric3}), 
\be
\ddot{h} + H\dot{h} = \frac{a^2}{\kappa^2}\left[\delta T^0_0-\delta T^i_i\right],
\ee
and averaging over the oscillating time derivatives $\ddot{\varphi}$, $\dot{\varphi}$ and $\delta_r$, one can combine Eqs. (\ref{edelta}) and 
(\ref{etheta}) 
into 
\be \label{edelta2}
\ddot{\delta} + H\dot{\delta} - \frac{a^2}{2\kappa^2} \rho_X \delta_X - C'\dot{\phi}\theta = [-C'k^2 + HC''\dot{\phi} + \ddot{C'} - 
                                \frac{a^2}{\kappa^2}V']\varphi
                              + \frac{a^2}{2\kappa^2}(\rho_c\delta_c+\rho_b\delta_b),
\ee
when $c_b^2 \approx 0$. Averaging out the small and oscillatory $\ddot{\varphi}$ and $\dot{\varphi}$ also from Eq.(\ref{klein}) yields
\be \label{phiper}
\varphi = -\frac{a^2\rho_X C'\delta_X + \frac{1}{2}\dot{\phi}\dot{h}}{a^2(V'' + \rho_X C'') + k^2}.
\ee
This approximation is in fact reasonably good for all scales. We have tested it by comparing the LHS and RHS of this expression 
evaluated by numerically integrating the full perturbation equations\footnote{In numerical calculations we use the public 
CAMB code\cite{Lewis:1999bs}, see http://camb.info/. For simplicity, we assume scale-invariant initial power spectrum, no reionization and three 
massless neutrino species. In our calculations, $h=0.71$. Normalization is such that the primordial curvature perturbation equals unity.}. Inserting 
the approximation in Eq.(\ref{edelta2}) and using Eq.(\ref{etheta}), 
we get in the small scale limit 
%\cite{Amendola:2003wa}  
\be \label{edelta3}
\ddot{\delta}_X + \left(H+C'\dot{\phi}\right)\dot{\delta}_X = \frac{3}{2}H^2\left[\Omega_X\left(1+2\frac{C'^2}{\kappa^2}\right)\delta_X + 
\Omega_c\delta_c + \Omega_b\delta_b\right].
\ee
Here we have used $C''=0$. During tracking era, $\Omega_X \sim a^{3[(1+w_B)(\frac{\beta}{\alpha}+1)-1]}$ vanishes to a good approximation. 
If we assume that then the noninteracting matter perturbation scales as $\delta_c \sim \tau^{m-2}$, we can use the tracker solution Eq.(\ref{tracker}) 
to find that 
\be \label{edelta4}
\ddot{\delta}_X + \left[1+\frac{3\beta}{\alpha}(1+w_B)\right]\frac{2}{(1+3w_B)\tau}\dot{\delta}_X = A\tau^m,
\ee
with the constant $A$ determined by the magnitude of noninteracting matter perturbations at a given time. The full solution to this equation is
\be
\delta_X = D + B\tau^{1-b} + \frac{A}{(b+m+1)(m+2)}\tau^{m+2}, 
\ee
where $B$ and $D$ are integration constants and $b=[2+6\beta(1+w_B)/\alpha]/(1+3w_B) \gg 1$. Thus the last term gives the dominant mode, which has the
same scaling as the noninteracting DM. Therefore the ratio $\delta_X/\delta_c$ is constant during tracking era. 
During matter domination, when $m=2$, this means that $\delta_X \sim a$, and $\delta_X/\delta_c = 3(1-3/\alpha^2)/(10+12\beta/\alpha) 
\ll 1$. 

\begin{figure}[h]
\begin{center}
\includegraphics[width=0.7\textwidth]{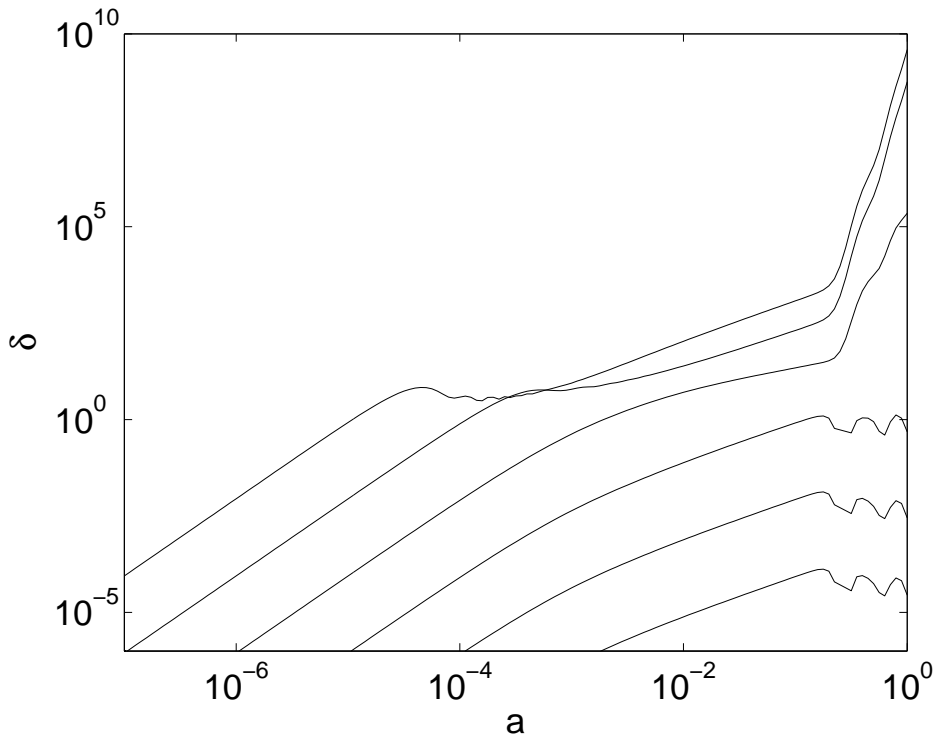} 
\caption{\label{deltas_i} Interacting DM overdensities in the model of section III. The slope $s$ of the curves in the logarithmic plot is 
$s \approx 2$, $1$, $11.2$ early in the radiation domination (when the scale is outside the horizon), during matter domination and during the 
transition to accelerating universe, respectively. The $k$-values from bottom to top are $k = 2.4 \cdot 10^{-6}$, $2.4 \cdot 10^{-5}$, $2.4 
\cdot 10^{-4}$, $0.0024$,
$0.024$ and $0.24$ Mpc$^{-1}$.}
\end{center}
\end{figure}

When the field reaches the minimum of its effective potential, it evolves with the rate\cite{Huey:2004qv}
\be \label{trans1}
\kappa\dot{\phi} = \frac{3}{\alpha+\beta}H.
\ee
The critical period we are interested in is the transition to the joint domination of quintessence and interacting matter that is ongoing today. The 
field derivative is then oscillating about the value given by Eq.(\ref{trans1}), to which it will asymptotically settle (see Fig.(\ref{eos_i})). Since 
$\Omega_X$ is now approaching $\Omega_c$, and $(1+C'^{2}) \gg 1$, we approximate Eq.(\ref{edelta3}) by dropping the source terms due to noninteracting 
matter. This approximation is well justified since, as it turns out, the RHS of Eq.(\ref{edelta3}) is soon dominated by the rapidly growing 
perturbation in the density of interacting DM.
We therefore write 
\be \label{edelta10}
\ddot{\delta}_X + \left(1+\frac{3\beta}{\alpha+\beta}\right)H\dot{\delta}_X = \frac{3}{2}H^2\Omega_X\left(1+2\beta^2\right)\delta_X.
\ee
To solve this equation analytically, we would need expressions for the background quantities $H$ (which follows from $w_B$) and $\Omega_X$. To that 
end we use the asymptotic value also for the latter  
\be \label{trans2}
w_B=-\beta/(\alpha+\beta), \quad \Omega_X = \alpha/(\alpha+\beta).
\ee
Inserting these in Eq.(\ref{edelta10}) yields the solutions
\be
\delta \sim a^n, \quad n= \frac{1}{4}\frac{(\alpha+10\beta)}{(\alpha+\beta)}
                 \left[-1 \pm \sqrt{1 + \frac{24\alpha(\alpha+\beta)(1+2\beta^2)}{(\alpha+10\beta)^2}}\right]. 
\ee
The exponent of the growing solution (corresponding to the positive sign) is large for typical values in this model. For $\alpha=5$ and $\beta=15$, $n 
\approx 
11.2$. This is in strong contrast to standard DM perturbations growth rate of which is slowing after the matter domination when $\delta_c \sim a$. 
Therefore 
\be
\theta_X \sim \dot{\delta}_c - \dot{\delta}_X \sim a^r, \quad r = n - \frac{2\beta-3\alpha}{2(\alpha+\beta)}, 
\ee
which is, consistently, the growing solution of Eq.(\ref{etheta}) in this limit. Numerical results are shown in Fig.(\ref{deltas_i})\footnote{Since
we have normalized the primordial curvature perturbation to unity, these perturbations in fact exit the linear regime at about $\delta \sim 10^5$. 
So these calculations are untrustworthy beyond those values, but that is unrelevant to our conclusions.}. 
There we see that the interacting DM density contrast adapts to this asymptotic scaling in fact as the transition to acceleration begins, 
although the solution Eqs.(\ref{trans1}),(\ref{trans2}) has not been reached. 

We conclude that when the energy in the $\chi$-matter is becoming a significant contributor to the background expansion, the perturbation
growth in this component is almost explosively sped up by the interaction pumping energy into it from the $\phi$-field. 
As seen from Eq.(\ref{phiper}), the DE is then inhomogenenous on small scales. It has been found also that in the nonlinear 
regime DE tends to clump with DM if coupled\cite{Mota:2004pa,Nunes:2004wn,vandeBruck:2005ii}. The large coupling 
for which $C'=\beta$ in this model results in a large effective gravitational constant in Eq.(\ref{edelta10}), $G^{eff} = (1+2\beta^2)G$. In earlier 
cosmology this is not as important, since then the standard DM overdensities constitute the dominant gravitational sources. There the modified Hubble 
friction term in Eq.(\ref{edelta3}) acts to reduce the relative magnitude of $\delta_X$ to $\delta_c$, but these evolve with the same rate. 
However, in the late universe the cosmic structure is drastically increased, except at the largest scales where we see only little oscillations. 
Consequently, the shape of the matter power spectrum in this model is highly incompatible with observations. Also, as the increase in structure will 
cause rapid deepening of the gravitational wells, the resulting CMB spectrum is modified by a strong late integrated Sachs-Wolfe effect due to time 
variation of the gravitational potentials. 
We have confirmed these considerations by numerical integration for several parameter values satisfying the relations given in the previous 
subsection. As a non-negligible amount of $\chi$-matter and a relatively strong coupling are required in this model, we feel it safe to state 
although having not performed an extensive exploration in the space of all cosmological parameters, that this model cannot be succesfully confronted 
with observations. 
%small scales the interaction pumping energy into $\chi$-matter will considerably speed up the growth of perturbations in this 
%component. In the synchronous gauge formalism this happens via the source term for the velocity perturbation in Eq.(\ref{etheta}). In the matter 
%overdensity evolution equation (\ref{edelta3}) the new force appears to work against the Hubble friction. When $C'\phi>H$, the net effect is to 
%enchance the growth. At redshifts $z \simeq \mathcal{O}(10)$, the density in $\chi$-matter has grown comparable to that in the standard CDM, but even 
%before that has $\rho\delta$ grown equal to $\rho_c\delta_c$. After that the structure in the universe increases drastically, except for the very 
%largest scales. Since $\nabla^2 \Psi \sim \delta$, gravitional wells grow rapidly for a large range of scales, which will be seen in the CMB spectrum 
%as an 
%integrated Sachs-Wolfe effect. We have confirmed these considerations by numerical integration, and found that for parameter values satisfying the 
%relations given in the previous subsection, the CMB and matter power spectra in this model are highly incompatible with observations. 
 
\section{Model with polynomial coupling}

\subsection{Background solution}

\begin{figure}[h]
\begin{center}
\includegraphics[width=0.7\textwidth]{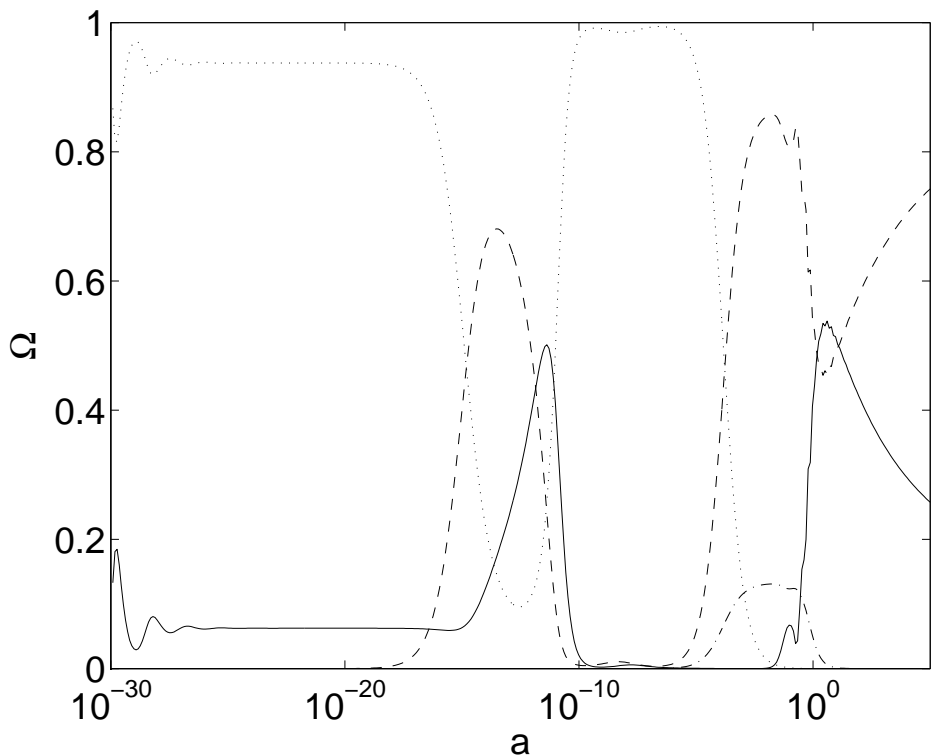} 
\caption{\label{omegas_d}Evolution of the fractional energy densities in the model of section IV. The solid line is $\Omega_\phi$, the dashed line 
$\Omega_X$. Dotted line is $\Omega_r$ and the dash-dotted line $\Omega_b$.}
\end{center}
\end{figure}

\begin{figure}[h]
\begin{center}
\includegraphics[width=0.45\textwidth]{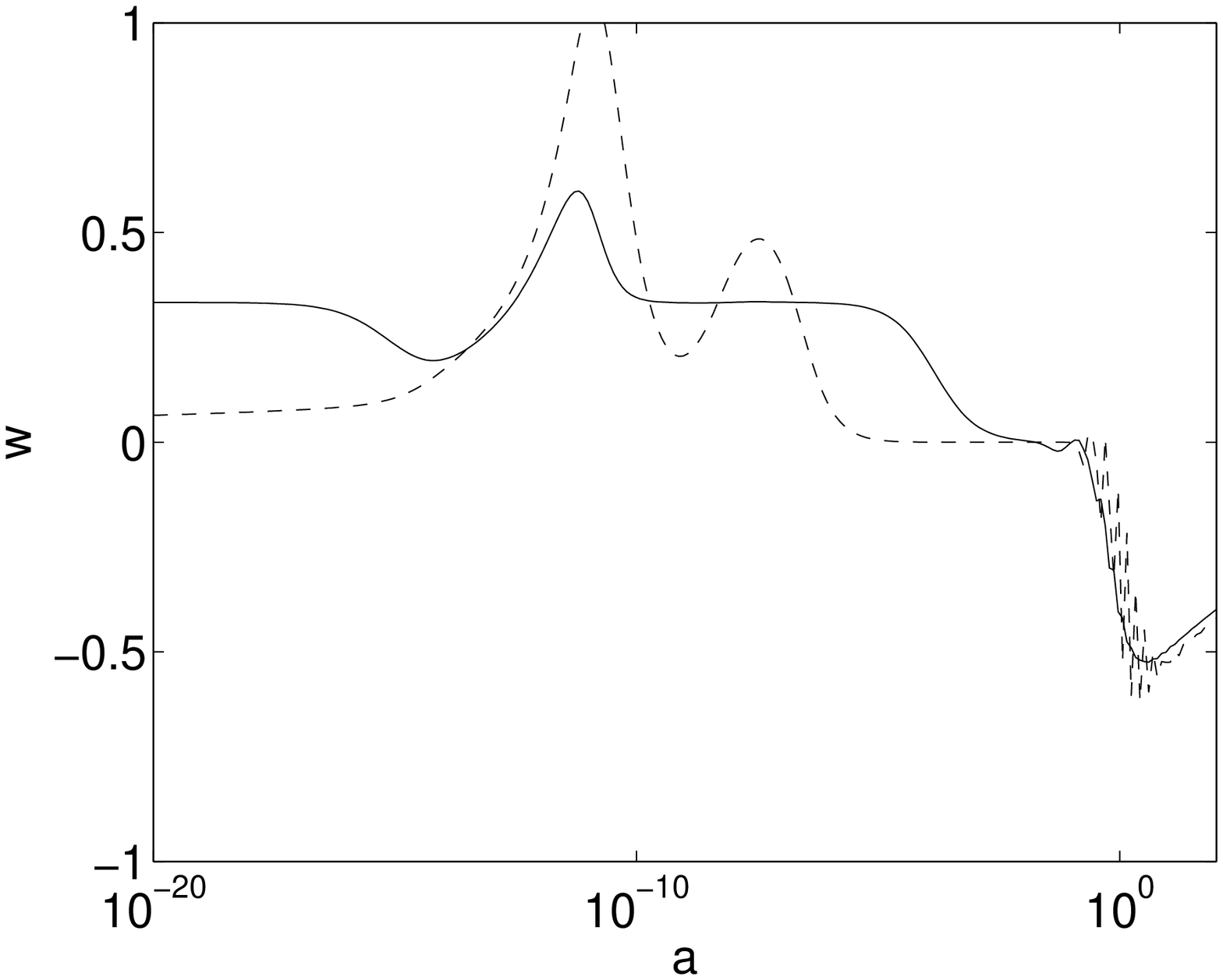} 
\includegraphics[width=0.45\textwidth]{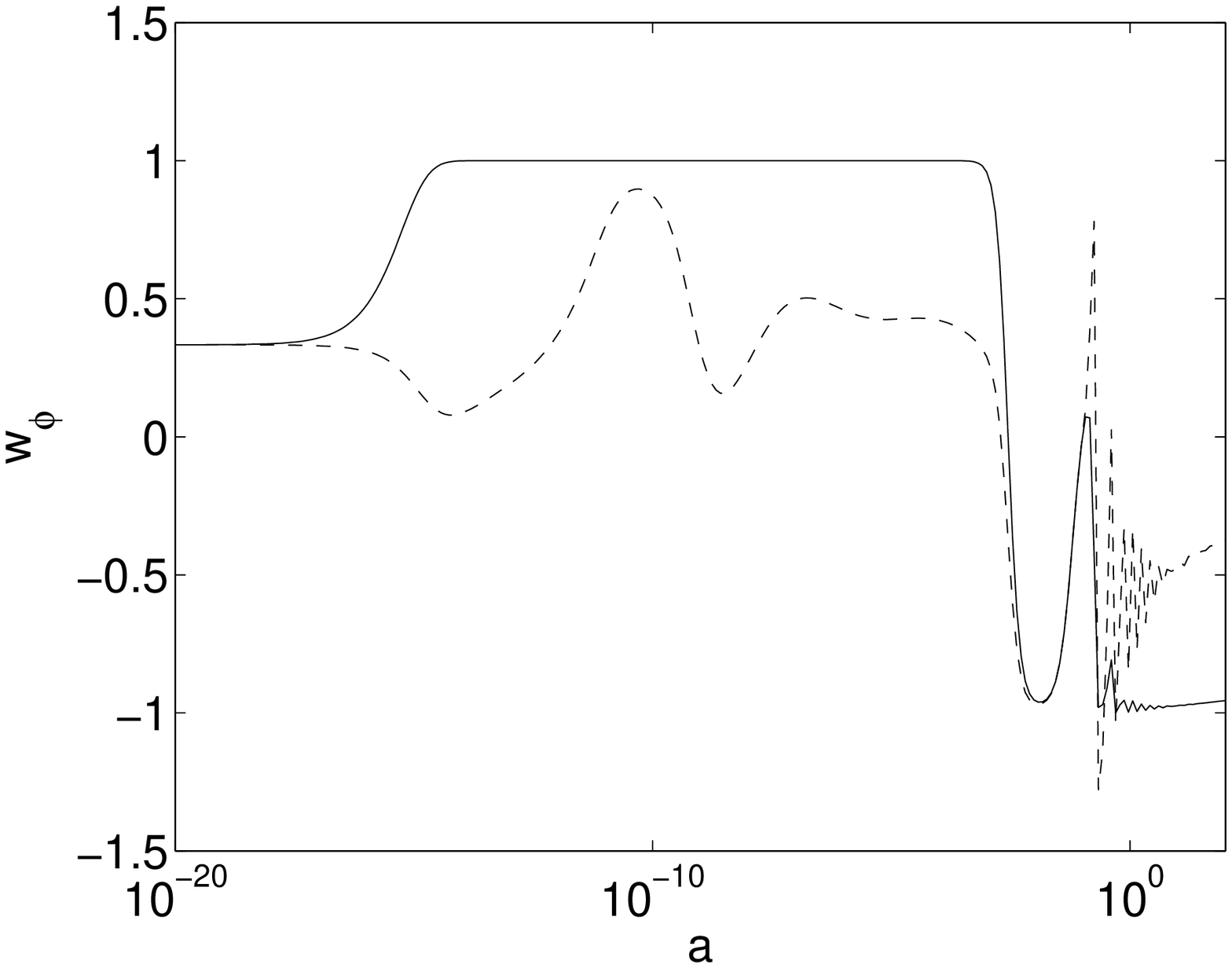} 
\caption{\label{eos}Left panel: Total equation of state $w$ (solid line) and the effective $w_X^{(eff)}$ (dashed line) for interacting DM in the 
model of section IV. Since $w<-1/3$ today, the universe accelerates. Right panel: Equation of state for the quintessence for the same model. Solid 
line is the formal $w_\phi$, and dashed line is the effective $w_\phi^{(eff)}$.}
\end{center}
\end{figure}

The model proposed by Bean and Magueijo\cite{Bean:2000zm} is another example of a modified exponential quintessence scenario. Again a
field is coupled to DM in such a way that quintessence is dominant today and driving the acceleration of the universe while the model 
still has its parameters of the order of the Planck scale and its outcome widely independent of the initial conditions for the field. This is achieved 
with 
only one DM species by using a non-exponential form for the coupling. To 
be definite, we have in the action Eq.(\ref{action}) 
\be
V(\phi) = V_0 e^{-\kappa\alpha\phi},
\quad
\gamma_\chi(\phi) = 1 + \lambda[\kappa(\phi-\phi_0)]^b.   
\ee
For numerical values we will use $\alpha = 8$, $b=8$, $\lambda=50$ and $\kappa\phi_0=32.5$. 
The background evolution is shown is Fig.(\ref{omegas_d}).
In this model the field leaves the standard exponential track long before matter domination. After initial transients, the field 
tracks the background radiation, until the coupling term pushes the field into kination. This lasts until the field reaches the minimum of its 
effective potential. Then it will at due time drive the acceleration. However, as the matter density eventually dilutes, the coupling force will 
fade, leading the field back to standard scaling in the asymptotic future. Thus matter domination is restored and the acceleration in this model 
is transient. We plot the effective equation of state parameter for the field $\phi$, as defined in Eq.(\ref{wpeff}), in the left panel of 
Fig.(\ref{eos}). In the right panel we plot the total equation of state for the universe together with the $w_X^{(eff)}$ as defined in 
Eq.(\ref{wxeff}). Since in this model $\rho_X$ governs the background expansion also in the matter dominated era, it is important that the coupling 
term has then negligible effect on the scaling of this density, since a significant deviation from $a \sim \tau^2$ would result in modified conformal 
distance to the last scattering surface, possibly dangerous to the compatibility with observations of the peaks in the fluctuation spectrum of the 
CMB radiation\cite{Hoffman:2003ru}. One also expects that then the growth rate of DM perturbations will not be much modified.
However, one would guess that the nonstandard evolution during transition to DE domination, seen in Figs.(\ref{omegas_d}) and (\ref{eos}), has 
some effects on the perturbation growth as was the case in the model of section III. Indeed these effects prove fatal to this model, as will be 
seen below.

\subsection{Perturbation evolution}

\begin{figure}[h]
\begin{center}
\includegraphics[width=0.7\textwidth]{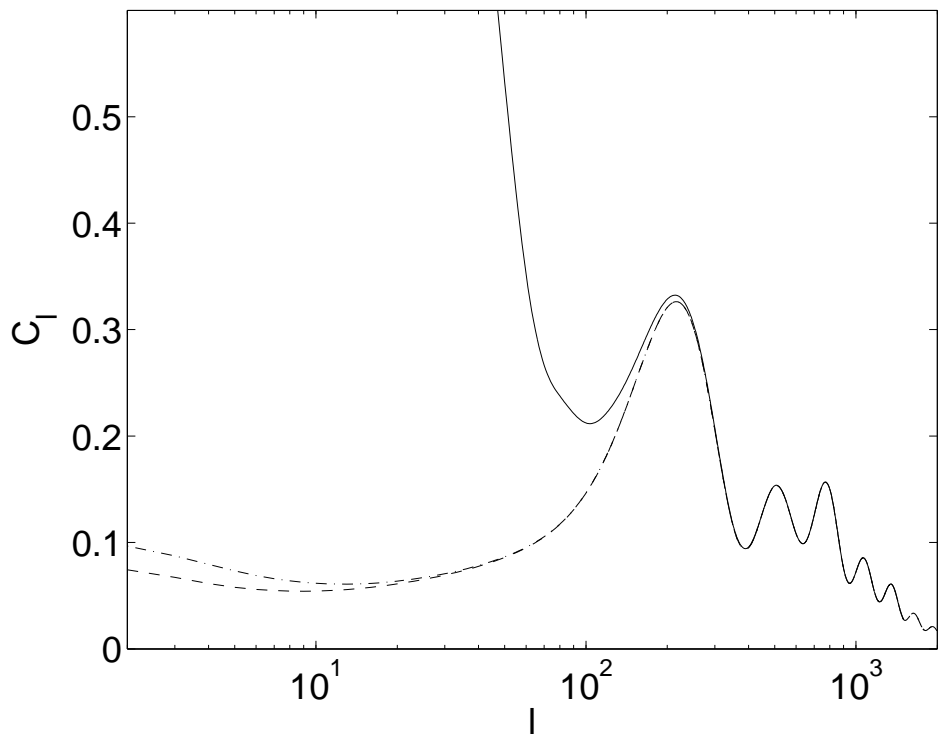} 
\caption{\label{spek} The CMB spectrum in the model of section IV (solid line). Dash-dotted line corresponds to suppressing the coupling of dark 
energy and DM perturbations (i.e. $\theta_\chi=0$), and dashed line corresponds to suppressing DE perturbations (i.e. $\varphi=0$).}
\end{center}
\end{figure}

The linearized Einstein equations have in fact been solved in this model\cite{Bean:2001ys}. However, this was not done taking fully into account 
the impact of the coupling to the matter perturbations, but instead by assuming that the fluctuations in $\rho_\chi$ decouple from those in 
$\phi$. This was realised by setting the DM velocity perturbation to zero, as conventional in the synchronous gauge. However, now the interaction
with DE will drive $\theta_\chi$ to evolve even if its initial value is chosen to be zero. One sees that the source term in the Euler 
equation Eq.(\ref{vel}) vanishes only in the absence of coupling or if the DE is perfectly smooth. Neither is the case now. We show the 
effect of physical assumptions in Fig.(\ref{spek}).

\begin{figure}[h]
\begin{center}
\includegraphics[width=0.45\textwidth]{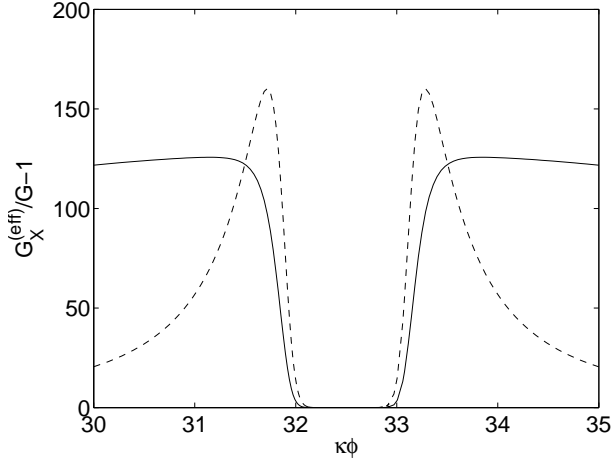} 
\caption{\label{coup} The deviation of the effective gravitational constant from the standard one, $G^{(eff)}/G-1$, as a function of the field 
$\kappa\phi$ for the model of section IV. The solid curve corresponds to the $G_\chi^{(eff)}$ in Eq.(\ref{edelta5}), and the dashed curve is $2C'^2$,  
corresponding to $G^{eff}_X$ as it appears in Eq.(\ref{edelta3}). For both definitions the deviation is small when $\phi \approx \phi_0$, but becomes 
very rapidly large when the field $\phi$ evolves to larger values.} 
\end{center}
\end{figure}

\begin{figure}[h]
\begin{center}
\includegraphics[width=0.7\textwidth]{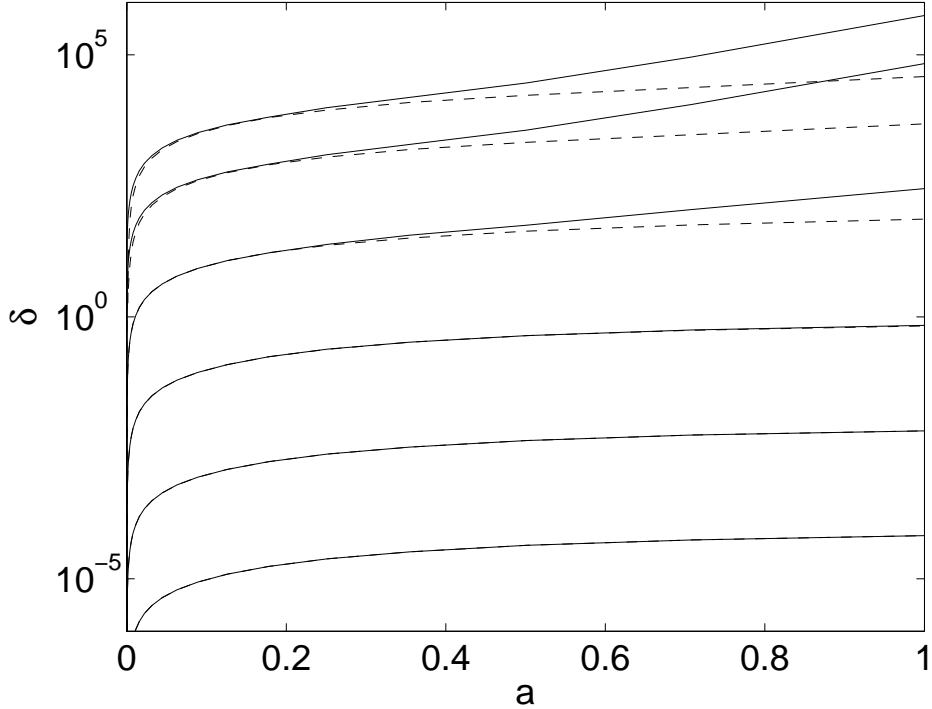} 
\caption{\label{deltas_d} DM overdensities (solid curves) in the model of section IV. Dashed curves are $\delta_b$. The $k$-values from 
bottom to top are $k = 2.4 \cdot 10^{-6}$, $2.4 \cdot 10^{-5}$, $2.4 \cdot 
10^{-4}$, $0.0024$, $0.024$ and $0.24$ Mpc$^{-1}$.}
\end{center}
\end{figure}

Since now $C''=0$, Eq.(\ref{edelta3}) is not valid. However, using the approximations that lead to that equation, we can derive the corresponding 
evolution equation for the quantity $\delta_\chi$. It should capture the qualitative behaviour of the DM perturbation, since $\delta_X=
\delta_\chi +C'\varphi \approx \delta_\chi$ for small scales. The equation we find reads
\be \label{edelta5}
\ddot{\delta}_\chi + \left(H-\frac{\lambda b y^{b-1}}{1+\lambda y^b}\kappa\dot{\phi}\right)\dot{\delta}_\chi = 
                \frac{3}{2}H^2\Omega_X\left[1+\frac{\lambda y^b(\lambda y^{b-2}(2b^2-y^2)-1)}{(1+\lambda y^b)^2}\right]\delta_\chi, 
\ee
where $y \equiv \kappa(\phi-\phi_0)$. The effective gravitational constant $G_\chi^{eff}$ can be thought to be the 
expression is square brackets times $G$. We plot $G^{(eff)}/G-1$ as a function of $\phi$ in Fig.(\ref{coup}) for a choice of parameters. 
The values of $G^{eff}$ becomes relevant at the matter dominated era, when the source term in the 
RHS of Eq.(\ref{edelta5}) is non-negligible. However, as the coupling has driven $\phi$ near to $\phi_0$, we have then $G_\chi^{eff} \approx G$ 
(for our example parameters, $-0.5<y<0$ from $a = 10^{-4}$ until $a = 0.1$). This changes when $\Omega_\phi$ climbs up, since then $\phi$ rolls 
forward. Similarly as in the previous case, $\delta_X$ is driven to rapid increase. We show its evolution in Fig.(\ref{deltas_d}), where we also 
plot the baryon overdensities $\delta_b$ for comparison. The boost experienced by $\delta_X$ is not as vast as we saw in the previous section, but still 
results in an order of magnitude or two too large $\delta_X$. Again we believe that notable improvement is not found by adjustments of parameters, but 
by revision of the scenario.


Finally, a remark about the adiabatic initial conditions we have used in the calculations. Adiabaticity of perturbations\cite{Liddle:2000cg} 
means that given any two superhorizon regions $A$ and $B$ in a perturbed universe at an instant $\tau$, $B$ is identical to $A$ at some other instant
$\tau + \delta \tau$. Evolution of the regions is the same, but slightly asynchronous. Then the ratio of fluctuation and 
time derivative of say a density field is common to all density fields. Thus we can define the entropy between species 
$i$ and $j$ as
\be \label{entropy}
S_{ij} \equiv \frac{\delta \rho_i}{\dot{\rho_i}} - \frac{\delta \rho_j}{\dot{\rho_j}} 
         =    \frac{\delta_i}{1+w_i^{(eff)}} - \frac{\delta_j}{1+w_j^{(eff)}}. 
\ee
The general adiabatic condition then dictates that all $S_{ij}$'s together with their derivatives vanish. Note that only in the absence of 
interactions does this condition coincide with the usually cited 
\be \label{entropy2}
\frac{\delta_i}{1+w_i} = \frac{\delta_j}{1+w_j}, \quad \theta_i = \theta_j. 
\ee 
There can be difference between these conditions, as seen for the cases we are now considering in Figs.(\ref{eos_i}) and (\ref{eos}). For a scale far 
outside the horizon and deep in the radiation dominated era, the growing adiabatic mode\cite{Ma:1995ey, Bucher:1999re} of the scalar field 
perturbation satisfies
\be
\varphi = \frac{3\phi^2-(\frac{1}{2}V'-\frac{1}{6}\gamma_\chi'\rho\chi)\dot{\phi} \tau
          +\frac{1}{2}\rho_\chi\left[\gamma_\chi''\dot{\phi}^2 - \gamma_\chi'a^2V'-\gamma_\chi'^2a^2\rho_\chi\right]\tau^2}
          {(6\dot{\phi}+\gamma_\chi'a^2\rho_\chi \tau)V' - \gamma_\chi''\rho_\chi \dot{\phi}^2\tau} \dot{\phi}Ak^2\tau^2,
\ee
\be
\dot{\varphi} =  \frac{(\frac{4}{3}\gamma_\chi'\rho_\chi a^2 V'+\frac{1}{2}a^2V'^2 -\frac{1}{2}\gamma_\chi''\rho_\chi\dot{\phi}^2)\tau
               + a^2\rho_\chi\left[-\frac{1}{2}\gamma_\chi''(V'+\frac{1}{3}\gamma_\chi'\rho_\chi) + \frac{1}{2}\gamma' a^2 V' +
                 \frac{4}{3} \gamma_\chi'^2 a^2\rho_\chi V'\right]\tau^2}
          {(6\dot{\phi}+\gamma_\chi'a^2\rho_\chi \tau)V' - \gamma_\chi''\rho_\chi \dot{\phi}^2\tau} \dot{\phi}Ak^2\tau^2.
\ee
Normalization $A=-1/2$ is convenient and sets comoving curvature perturbation equal to unity. Fortunately the above expressions can in many 
occasions be simplified by dropping the square bracket terms, but we have kept them here for the sake of completeness. 
The departure of the usual adiabatic condition Eq.(\ref{entropy2}) from 
the generalized one has been pointed out also in Refs.\cite{Malik:2002jb,Malik:2004tf}. 
If the 
field is a tracker, the initial conditions may indeed be irrelevant also for the perturbed field. However, for minimally coupled fields it has been 
shown that in some instances this is not the case\cite{Kawasaki:2001nx,Abramo:2001mv}, while isocurvature perturbations in 
coupled quintessence remain to be systematically studied.   

\section{Discussion}

Interestingly, in the two models we have investigated, the growth of DM perturbations is enchanced, whereas in minimally coupled models 
the late time effect of DE is the opposite. This may be understood by comparing to the case where the interacting DE and DM satisfy the adiabatic 
relation $S_{\phi \chi}=\dot{S}_{\phi \chi}=0$, where $S_{\phi \chi}$ is defined by Eq.(\ref{entropy}). Then the system of these components 
can be treated as a unified fluid without intrinsic entropy. It is known that the effective sound speed squared $c_S^2 \equiv \dot{p}/\dot{\rho}$ of 
such a fluid governs the late dynamics of the perturbations\cite{Reis:2005wz, Koivisto:2004ne}. If $c_S^2>0$, they are driven to oscillate, if 
$c_S^2<0$ they tend to blow up. This is the work of the pressure gradient sourcing the evolution equation one can derive for the unified fluid 
density contrast,
\be \label{udeltae}
\ddot{\delta}+[1-3(2w-c_S^2)]H\dot{\delta}-\frac{3}{2}(1-6c_S^2+8w-3w^2)H^2\delta = -k^2 \frac{\delta p}{\delta \rho} \delta.
\ee
This equation is exact (to linear order, in the comoving gauge). As $w$ and $c_S^2$ are given by the background expansion, one needs only information 
of the pressure perturbation to solve the structure in the universe. In the adiabatic case, $\delta p/\delta \rho = c_S^2$, but in the scalar 
field models considered here this assumption is not valid. However, using equations from subsection IIB, the small scale limit of the pressure 
gradient in the late universe is readily found. It is 
\be \label{gradient}
-k^2\frac{\delta p}{\delta \rho} = -k^2\frac{\frac{1}{a^2}\dot{\phi}\dot{\varphi}-V'(\phi)\varphi}{\rho_\chi\delta_\chi -2 V'(\phi)\varphi}
                                 \rightarrow -a^2 C'(\phi) V'(\phi),
\ee
(also in the comoving gauge) and turns out positive for the models of section III and IV. Thus one may regard the enchanced small scale growth rate 
there as a consequence of negative effective (non-adiabatic) sound speed squared. Now the pressure gradient approaches a $k$-independent expression and the 
growth rate saturates at a given value when $k$ is increased beyond some critical scale, unlike in the adiabatic case. Note also that the total 
density perturbation $\delta$ in Eq.(\ref{udeltae}),
\be
\delta = \frac{\rho_\chi \gamma'_\chi(\phi)\delta_\chi + [\rho_\chi \gamma'(\phi) + V'(\phi)]\varphi + \frac{1}{a^2}\dot{\phi}\dot{\varphi}}
              {\gamma_\chi(\phi)\rho_\chi + \frac{1}{a^2}\dot{\phi}^2 + V(\phi)}, 
\ee
is directly relevant to the integrated Sachs-Wolfe effect in the CMB fluctuation spectrum.

Before concluding, we use the opportunity to comment on yet another model with DE-DM interaction\cite{Chimento:2003ie}. There DE with a constant 
equation of state parameter $w_\phi$ is coupled to DM in such a way that 
\be \label{cont1}
\dot{\rho}_\phi + 3H\rho_\phi = 3H \lambda^2 (\rho_X + \rho_\phi),
\ee
\be \label{cont2}
\dot{\rho}_X + 3H\rho_X = -3H \lambda^2 (\rho_X + \rho_\phi),
\ee
where $\lambda^2$ is a constant measuring the intensity of the interaction. It might be difficult to find a field theory underlying this 
phenomenological model, but that would be necessary to be able to derive the corresponding equations to the linear order. Otherwise they must be put 
in by hand. 
%The simplest possibility would be to assume no entropy as defined by Eq.(\ref{entropy}) between these components, 
%but that would exclude the case of a scalar field and probably result in unwanted oscillations or blow-ups in the matter power 
%spectrum\cite{Koivisto:2004ne, Reis:2005wz}. 
The linear DE continuity equations presented in the recent paper\cite{Olivares:2005tb} imply that there
\be \label{div1}
\nabla_\mu {T^\mu_0}^{(\phi)} = 3H\rho_\phi \lambda^2(\delta_\phi + r\delta_X),
\ee
\be \label{div2}
\nabla_\mu {T^\mu_i}^{(\phi)} = [3H\rho_\phi \lambda^2(1+r)\theta_\phi]_{,i},
\ee      
where $r \equiv \rho_\chi/\rho_\phi$. Except perhaps for the sign of the $0$-component of the divergence\footnote{The sign is different from the 
zero-order divergence. Possibly this is a misprint.}, these expressions are not unreasonable, but do not follow unambiguously from Eqs.(\ref{cont1}) 
and (\ref{cont2}). However, once the divergence of the DE stress tensor is given, the interacting DM 
continuity equations are uniquely fixed. When Eqs.(\ref{div1}) and (\ref{div2}) hold for the DE and the DM Lagrangian is effectively 
of the form (\ref{lagran}), the Bianchi identities determine that
\be
\dot{\delta}_X=-\theta_X-\frac{1}{2}\dot{h}-3H\lambda^2(\frac{1}{r}\delta_\phi+\delta_X),
\ee 
\be \label{joku}
\dot{\theta}_X=-H\theta_X+3H\lambda^2(1+\frac{1}{r})\theta_\phi.
\ee
From these we recover the equations used in Ref.\cite{Olivares:2005tb} only by setting $\theta=0$ in the first equation and forgetting about the 
second one. But as detailed in the previous section, such approximation can leave crucial features out from the evolution of perturbations. Therefore 
we must regard that the results of the data analysis in Ref.\cite{Olivares:2005tb}, where it is announced that observations prefer interacting DE 
models, call for reinvestigation. 

%To conclude, we considered the linear growth rate in models with a DE-DM interaction. In particular, we studied two 
%such models and found both of them featuring very large growth of density perturbations in the late universe. This was a consequence of the momentum 
%transfer from the perturbed DE field to the coupled DM, and could be understood by finding the effective strenght of gravitation experienced by the 
%interacting DM, or the total pressure gradient of the cosmic fluid in the small wavelenght limit. 

As mentioned in the beginning of this paper, the origin of most of the energy density in the universe is unknown. Suggestions can be 
based on speculations of string theory, but this is tentative at the present. However, some inventive mechanism, as the ones considered here, is 
necessary if some light is required to be shed on the coincidence problem. Moreover, as the results of this paper demonstrate, considerations of 
cosmological structure provide powerful means to constrain such mechanisms.

\acknowledgments{The author acknowledges useful discussions with Hannu Kurki-Suonio, David Mota and Finn Ravndal. This work was supported in part by 
NorFA, Waldemar von Frenckels Stiftelse and Emil Aaltosen S\"a\"ati\"o. 

\bibliography{refs}

\end{document}